\documentstyle[sprocl]{article}

\def\sst{\scriptscriptstyle}

\begin{document}

\begin{titlepage}

\title{
\vspace*{-1.8cm}
\begin{flushright}
{\normalsize DO--TH 00/01\\LNF-00/001 (P)\\[2.0cm]}
\end{flushright}
{\Large \bf Analyzing \boldmath $\varepsilon'/\varepsilon$ 
\unboldmath in the \boldmath $1/N_c$ \unboldmath Expansion}
\vspace*{0.8cm}
}
\author{
\large
T.\ Hambye$^{1\,}$\footnote{\small E-mail: hambye@lnf.infn.it},
G.O.\ K\"ohler$^{2\,}$\footnote{\small E-mail:
koehler@doom.physik.uni-dortmund.de}, 
E.A.\ Paschos$^{2\,}$\footnote{\small E-mail:                     
paschos@hal1.physik.uni-dortmund.de},
and P.H.\ Soldan$^{2\,}$\footnote{\small E-mail: 
soldan@doom.physik.uni-dortmund.de}\\[1cm]
\normalsize 1: {\it INFN - Laboratori Nazionali di Frascati,
P.O. Box 13,\\ I-00044 Frascati, Italy}\\[2mm]
\normalsize 2: {\it Institut f\"ur Physik, Universit\"at Dortmund
D-44221 Dortmund, Germany}\\[2cm]
}
\date{}

\vspace*{5mm}

\maketitle

\begin{abstract}
We present a recent analysis of $\varepsilon'/\varepsilon$ in the 
$1/N_c$ expansion. We show that the $1/N_c$ corrections to the matrix 
element of $Q_6$ are large and positive, indicating a $\Delta I=1/2$ 
enhancement similar to the one of $Q_1$ and $Q_2$ which dominate the 
CP conserving amplitude. This enhances the CP ratio and can bring the 
standard model prediction close to the measured value for central 
values of the parameters.
\end{abstract}

\vspace*{1.8cm}
\noindent
Talk presented by P.H.\ Soldan at The 3$^{\mbox{rd}}$ International 
Conference on ${\cal B}$ Physics and ${\cal CP}$ Violation, Taipei, 
Taiwan, December 3 - 7, 1999.

\end{titlepage}

\setcounter{footnote}{0}


\title{ANALYZING \boldmath $\varepsilon'/\varepsilon$ \unboldmath 
IN THE \boldmath $1/N_c$ \unboldmath EXPANSION\footnote{Talk 
presented by P.H.\ Soldan.}}

\author{T.~HAMBYE}
\address{INFN - Laboratori Nazionali di Frascati,I-00044 Frascati, Italy}
\author{G.O.~K\"OHLER, E.A.~PASCHOS, P.H.~SOLDAN}
\address{Institut f\"ur Physik, Universit\"at Dortmund,
D-44221 Dortmund, Germany}

\maketitle\abstracts{We present a recent analysis of $\varepsilon'/
\varepsilon$ in the $1/N_c$ expansion. We show that the $1/N_c$ 
corrections to the matrix element of $Q_6$ are large and positive, 
indicating a $\Delta I=1/2$ enhancement similar to the one of $Q_1$ 
and $Q_2$ which dominate the CP conserving amplitude. This enhances 
the CP ratio and can bring the standard model prediction close to 
the measured value for central values of the parameters.}

\section{Introduction}
Direct CP violation in $K\rightarrow\pi\pi$ decays was recently
observed by the KTeV and NA48 collaborations.\cite{ktev,fanti} The 
present world average \cite{fanti} for the parameter $\varepsilon'/
\varepsilon$ is $\mbox{Re}\,\varepsilon'/\varepsilon\,=\,(21.2 \pm 
4.6)\cdot 10^{-4}$. In the standard model CP violation originates 
in the CKM phase, and direct CP violation is governed by loop diagrams 
of the penguin type. The main source of uncertainty in the calculation 
of $\varepsilon'/\varepsilon$ is the QCD non-perturbative contribution 
related to the hadronic nature of the $K\rightarrow\pi\pi$ decay. Using 
the $\Delta S=1$ effective hamiltonian,
\begin{equation}
{\cal H}_{ef\hspace{-0.5mm}f}^{\sst\Delta S=1}=\frac{G_F}{\sqrt{2}}
\;\lambda_u\sum_{i=1}^8 c_i(\mu)\,Q_i(\mu)\hspace{1cm}(\mu < m_c)\,,
\label{ham}
\end{equation}
the non-perturbative contribution, contained in the hadronic matrix 
elements of the four-quark operators $Q_i$, can be separated from the 
perturbative Wilson coefficients $c_i(\mu)=z_i(\mu)+\tau y_i(\mu)$ 
(with $\tau=-\lambda_t/\lambda_u$ and $\lambda_q=V_{qs}^*\,V_{qd}^{}$). 
Introducing $\langle Q_i\rangle_I\equiv\langle (\pi\pi)_I|Q_i|K\rangle$, 
the CP ratio can be written as
\begin{equation}
\frac{\varepsilon'}{\varepsilon}\,=\,\frac{G_F}{2}
\frac{\omega\,\mbox{ Im}\lambda_t}{|\varepsilon|\,\mbox{Re}A_0}
\left[\,\Big|\sum_i\,y_i\,\langle Q_i\rangle_0\Big|\,
\Big(1-\Omega_{\eta+\eta'}\Big)\,\,-\,\frac{1}{\omega}
\Big|\sum_i\,y_i\,\langle Q_i\rangle_2\Big|\,\right].
\label{epspsm}
\end{equation}
$\omega=$Re$A_0/$Re$A_2=22.2$ is the ratio of the CP conserving $K
\rightarrow\pi\pi$ isospin amplitudes; $\Omega_{\eta+\eta'}$ encodes
the effect of the isospin breaking in the quark masses.\cite{ecker} 
$\varepsilon'/\varepsilon$ is dominated by $\langle Q_6\rangle_0$ and 
$\langle Q_8\rangle_2$ which cannot be fixed from the CP conserving 
data.\cite{BJM,bosch} Beside the theoretical uncertainties coming from 
the calculation of the $\langle Q_i\rangle_I$ and of $\Omega_{\eta
+\eta'}$, the analysis of the CP ratio suffers from the uncertainties 
on the values of various input parameters, in particular of the CKM 
phase in Im$\lambda_t$, of $\Lambda_{\mbox{\tiny QCD}}\equiv
\Lambda^{(4)}_{\overline{\mbox{\tiny MS}}}$, and of the strange 
quark mass. 

To calculate the hadronic matrix elements we start from the effective
chiral lagrangian for pseudoscalar mesons which involves an expansion 
in momenta where terms up to ${\cal O}(p^4)$ are included.\cite{GaL} 
The method we use is the $1/N_c$ expansion.\cite{tt,BBG2} In this 
approach, we expand the matrix elements in powers of the momenta and 
of $1/N_c$. For the $1/N_c$ corrections we calculated chiral loops as 
described in refs.~\cite{HKPSB,hks}. Especially important to this analysis 
are the non-factorizable corrections, which are UV divergent and must 
be matched to the short-distance part. They are regularized by a 
finite cutoff $\Lambda_c$ which is identified with the short-distance 
renormalization scale. The definition of the momenta in the loop 
diagrams, which are not momentum translation invariant, is discussed 
in detail in ref.~\cite{HKPSB}. Other recent work on matrix elements 
in the $1/N_c$ approach can be found in refs.~\cite{BP,kpr}. 

For the Wilson coefficients we use the leading logarithmic and the 
next-to-leading logarithmic values.\cite{BJM} The absence of any 
reference to the renormalization scheme in the low-energy calculation, 
at this stage, prevents a complete matching at the next-to-leading 
order.\cite{ab98} Nevertheless, a comparison of the numerical results 
obtained from the LO and NLO coefficients is useful as regards estimating 
the uncertainties and testing the validity of perturbation theory.

\section{Analysis of \boldmath $\varepsilon'/\varepsilon$\unboldmath}

Analytical formulas for all matrix elements, at next-to-leading order in 
the twofold expansion in powers of momenta and of $1/N_c$, are given in 
refs.~\cite{HKPSB,hks}. In the pseudoscalar approximation, the matching 
has to be done below 1\,GeV. Varying $\Lambda_c$ between 600 and 900\,MeV, 
the bag factors $B_1^{(1/2)}$ and $B_2^{(1/2)}$ take the values 
$8.2-14.2$ and $2.9-4.6$; quadratic terms in $\langle Q_1\rangle_0$ 
and $\langle Q_2\rangle_0$ produce a large enhancement which brings 
the $\Delta I=1/2$ amplitude in agreement with the data.\cite{hks} 
Corrections beyond the chiral limit were found to be small.

For $\langle Q_6\rangle_0$ and $\langle Q_8\rangle_2$ the leading
non-factorizable loop corrections, which are of ${\cal O}(p^0/N_c)$, 
are only logarithmically divergent.\cite{HKPSB} Including terms of 
${\cal O}(p^0)$, ${\cal O}(p^2)$, and ${\cal O}(p^0/N_c)$, $B_6^{(1/2)}$ 
and $B_8^{(3/2)}$ take the values $1.10-0.72$ and $0.64-0.42$. As a 
result the experimental range for $\varepsilon'/\varepsilon$ can be 
accommodated in the standard model only if there is a conspiracy of 
the input parameters.\footnote{For supersymmetric contributions to 
$\varepsilon'/\varepsilon$ see ref.~\cite{susy} and references therein.} 
However, since the leading ${\cal O}(p^0)$ contribution vanishes for 
$Q_6$, corrections from higher order terms beyond the ${\cal O}(p^2)$ 
and ${\cal O}(p^0/N_c)$ are expected to be large. In ref.~\cite{HKPS} 
we investigated the ${\cal O}(p^2/N_c)$ contribution, i.e., the $1/N_c$ 
correction at the next order in the chiral expansion, because it brings 
about, for the first time, quadratic corrections on the cutoff. 
From counting arguments and more generally from the fact that the 
chiral limit is assumed to be reliable, the quadratic terms (which 
are not chirally suppressed) are expected to be dominant. It is still 
desirable to check that explicitly by calculating the corrections 
beyond the chiral limit, from logarithms and finite terms, as done 
for $Q_1$ and $Q_2$. Numerically, we observe a large positive correction 
from the quadratic term in $\langle Q_6\rangle_0$. This point was 
already emphasized in ref.~\cite{orsay}. The slope of the correction 
is qualitatively consistent and welcome since it compensates for the 
logarithmic decrease at ${\cal O}(p^0/N_c)$. Varying $\Lambda_c$ 
between 600 and 900\,MeV, the $B_6^{(1/2)}$ factor takes the values 
$1.50-1.62$. $Q_6$ is a $\Delta I=1/2$ operator, and the enhancement 
of $\langle Q_6\rangle_0$ indicates that at the level of the $1/N_c$ 
corrections the dynamics of the $\Delta I=1/2$ rule applies to $Q_6$ 
as to $Q_1$ and $Q_2$. 

Using the quoted values for $B_6^{(1/2)}$ together with the full
leading plus next-to-leading order $B$ factors for the remaining
operators \cite{HKPS} we calculated $\varepsilon'/\varepsilon$.
The results for the three sets of Wilson coefficients LO, NDR, 
and HV and for $\Lambda_c$ between 600 and $900\,\mbox{MeV}$ are 
given in Tab.~\ref{tab1}. The numbers are close to the measured 
value for central values of the parameters (first column). They are 
obtained by assuming zero phases from final state interactions. This 
approximation is very close to the results we would get if we used 
the small imaginary part obtained at the one-loop level.\cite{HKPS} 
\noindent
\begin{table}[t]
\caption{Numerical values for $\varepsilon'/ \varepsilon$ (in units of
$10^{-4}$) as explained in the text.
\label{tab1}}
\begin{eqnarray*}
\begin{array}{|c|c|c|}\hline
\rule{0cm}{5mm}
\mbox{LO}
&\,\,\,14.8 \,\,\leq\,\,\varepsilon'/\varepsilon\,\,
\leq\,\,19.4 \,\,\,
& \,\,\,6.1 \,\,\leq\,\,\varepsilon'/\varepsilon\,\,
\leq\,\,48.5 \,\,\,\\[0.2mm]
\mbox{\,\,NDR\,\,}
&\,\,\,12.5 \,\,\leq\,\,\varepsilon'/\varepsilon\,\,
\leq\,\,18.3 \,\,\,
& \,\,\,5.2 \,\,\leq\,\,\varepsilon'/\varepsilon\,\,
\leq\,\,49.8 \,\,\, \\[0.2mm]
\mbox{HV}
& \,\,\,7.0\,\,\,\,\leq\,\,\varepsilon'/\varepsilon\,\,
\leq\,\,14.9 \,\,\,
& \,\,\,2.2\,\,\leq\,\,\varepsilon'/\varepsilon\,\,
\leq\,\,38.5 \,\,\, \\[1mm]
\hline    
\end{array}
\end{eqnarray*}
\end{table}

Performing a scanning of the parameters [$125\,\mbox{MeV}\leq m_s(1\,
\mbox{GeV})\leq 175$ $\mbox{MeV}$, $0.15\leq\Omega_{\eta+\eta'}\leq 0.35$, 
$1.04\cdot 10^{-4}\leq\mbox{Im}\lambda_t\leq 1.63\cdot 10^{-4}$, and $245
\,\mbox{MeV}\leq\Lambda_{\mbox{\tiny QCD}}\leq 405\,\mbox{MeV}$] we obtain 
the numbers in the second column of Tab.~\ref{tab1}. They can be compared 
with the results of refs.~\cite{bosch,BEF,bel,CM,blum}. The values of 
$B_6^{(1/2)}$ can also be compared with ref.~\cite{BP} and those of 
$B_8^{(3/2)}$ with refs.~\cite{kpr,don}. The large ranges reported in the 
table can be traced back to the large ranges of the input parameters. This
can be seen by comparing them with the relatively narrow ranges obtained for 
central values of the parameters. The parameters, to a large extent, act 
multiplicatively, and the large range for $\varepsilon'/\varepsilon$ is due 
to the fact that the central value(s) for the ratio are enhanced roughly by 
a factor of two compared to the results obtained with $B$ factors for $Q_6$ 
and $Q_8$ close to the VSA. More accurate information on the parameters, 
from theory and experiment, will restrict the values for $\varepsilon'/
\varepsilon$. 

To estimate the uncertainties due to higher order final state interactions 
we also calculated $\varepsilon'/\varepsilon$ using the real part of the 
matrix elements and the phenomenological values of the phases \cite{phases}, 
$\delta_0=(34.2\pm 2.2)^\circ$ and $\delta_2=(-6.9 \pm 0.2)^\circ$, i.e., 
we replaced $|\sum_i y_i\langle Q_i\rangle_I|$ in Eq.~(\ref{epspsm}) by 
$\sum_i y_i\mbox{Re}\langle Q_i\rangle_I/\cos\delta_I$. The corresponding
results are given in Tab.~\ref{tab2}. They are enhanced by $\sim 25\,\%$ 
compared to the numbers in Tab.~\ref{tab1}. We would like to emphasize that 
this $\sim 25\,\%$ uncertainty should be taken into account by any analysis 
which either does not include final state interactions or cannot reproduce 
the numerical values of the phases. To reduce the uncertainties in the 
$1/N_c$ approach it would be interesting to investigate the two-loop 
imaginary part and/or to combine our calculation with a dispersive 
calculation along the lines of refs.~\cite{pich,eap,truong}. In order to 
reduce the scheme dependence in the result, appropriate subtractions would 
be necessary (see refs.~\cite{BP,BB}). Finally, it is reasonable to assume 
that the effect of the pseudoscalar mesons is the most important one. 
Nevertheless, incorporating the vector mesons and higher resonances would 
be desirable in order to improve the treatment of the intermediate region 
around the rho mass and to show explicitly that the large enhancement we 
find at low energy in the treatment of the pseudoscalars remains valid up 
to the scale $\sim m_c$, where the matching with the short-distance part 
can be done more safely. 
\noindent
\begin{table}[t]
\caption{Same as in Tab.~\ref{tab1}, but now with the phenomenological 
values for the phases as explained in the text.
\label{tab2}}
\vspace*{-1.8mm}
\begin{eqnarray*}
\begin{array}{|c|c|c|}\hline
\rule{0cm}{5mm}
\mbox{LO}
&\,\,\,19.5 \,\,\leq\,\,\varepsilon'/\varepsilon\,\,
\leq\,\,24.7 \,\,\,
& \,\,\,8.0 \,\,\leq\,\,\varepsilon'/\varepsilon\,\,
\leq\,\,62.1 \,\,\,\\[0.2mm]
\mbox{\,\,NDR\,\,}
&\,\,\,16.1 \,\,\leq\,\,\varepsilon'/\varepsilon\,\,
\leq\,\,23.4 \,\,\,
& \,\,\,6.8 \,\,\leq\,\,\varepsilon'/\varepsilon\,\,
\leq\,\,63.9 \,\,\, \\[0.2mm]
\mbox{HV}
& \,\,\,9.3\,\,\,\,\leq\,\,\varepsilon'/\varepsilon\,\,
\leq\,\,19.3 \,\,\,
& \,\,\,2.8\,\,\leq\,\,\varepsilon'/\varepsilon\,\,
\leq\,\,49.8 \,\,\, \\[1mm]
\hline    
\end{array}
\end{eqnarray*}
\vspace*{-1.9mm}
\end{table}
\section*{Acknowledgments}
This work was supported by BMBF, 057D093P(7), Bonn, FRG, and DFG Antrag 
PA-10-1. T.H.~acknowledges support from EEC, TMR-CT980169.

\end{document}